\documentclass[%
 reprint,
 pra,
 floatfix,
 amsmath,amssymb,
 aps, twocolumn,
 superscriptaddress,
 longbibliography,
]{revtex4-2}

\usepackage{graphicx}
\usepackage{xcolor}
\usepackage{amsmath}
\usepackage{cleveref}
\usepackage{array,multirow}
\usepackage{xspace}

\newcommand{\rA}{A\xspace} %488
\newcommand{\rB}{B\xspace} %491

\allowdisplaybreaks[2]

\newcommand{\mpik}{\affiliation{Max Planck Institute for Nuclear Physics, 69117 Heidelberg, Germany}}
\newcommand{\hijena}{\affiliation{Helmholtz Institute Jena, 07743 Jena, Germany}}
\newcommand{\fsujena}{\affiliation{Friedrich-Schiller-Universit\"at Jena, 07743 Jena, Germany}}
\newcommand{\gsi}{\affiliation{GSI Helmholtzzentrum f\"ur Schwerionenforschung,  64291 Darmstadt, Germany}}
\newcommand{\desy}{\affiliation{Deutsches Elektronen-Synchrotron DESY, 22607 Hamburg, Germany}}
\newcommand{\cui}{\affiliation{The Hamburg Centre for Ultrafast Imaging CUI, 22761 Hamburg, Germany}}
\newcommand{\alufreiburg}{\affiliation{Albert-Ludwigs University of Freiburg,  79104 Freiburg, Germany}}
\newcommand{\euxfel}{\affiliation{European X-Ray Free-Electron Laser Facility, 22869 Schenefeld, Germany}}

\begin{document}

\title{Single-shot sorting of M\"ossbauer time-domain data at X-ray free-electron lasers}

\author{Miriam Gerharz}
\mpik

\author{Willi Hippler}
\hijena
\gsi

\author{Berit Marx-Glowna}
\hijena
\gsi

\author{Sakshath Sadashivaiah}
\hijena
\gsi

\author{Kai S. Schulze} % nur bei p2778 nicht p3334 dabei
\fsujena

\author{Ingo Uschmann}
\fsujena

\author{Robert Loetzsch}
\hijena
\gsi
\fsujena

\author{Kai Schlage}
\desy 

\author{Sven Velten}
\desy 
\cui

\author{Dominik Lentrodt}
\alufreiburg

\author{Lukas Wolff}  
\mpik

\author{Olaf Leupold}
\desy 

\author{Ilya Sergeev}
\desy 

\author{Hans-Christian Wille}
\desy 

\author{Cornelius Strohm}
\desy

\author{Marc Guetg}
\desy

\author{Shan Liu}
\desy

\author{Gianluca Aldo Geloni}
\euxfel

\author{Ulrike Boesenberg}
\euxfel

\author{J\"org Hallmann}
\euxfel

\author{Alexey Zozulya}
\euxfel

\author{Jan-Etienne Pudell}
\euxfel

\author{Angel Rodriguez-Fernandez}
\euxfel

\author{Mohamed Youssef}
\euxfel

\author{Anders Madsen}
\euxfel

\author{Lars Bocklage}
\desy 
\cui

\author{Gerhard G. Paulus}
\fsujena

\author{Christoph H. Keitel}
\mpik

\author{Thomas Pfeifer}
\mpik

\author{Ralf Röhlsberger}
\hijena
\gsi 
\fsujena 
\desy
\cui

\author{J\"org Evers${}^*$}
\mpik

\date{\today}

%%%%%%%%%% ABSTRACT %%%%%%%%%%%%%%%%%%%%%%
 
\begin{abstract}
Mössbauer spectroscopy is widely used to study structure and dynamics of matter with remarkably high energy resolution, provided by the narrow nuclear resonance line widths. However, the narrow width implies low count rates, such that experiments commonly average over extended measurement times or many x-ray pulses (``shots''). This averaging  impedes the study of non-equilibrium phenomena.
It has been suggested that X-ray free-electron lasers (XFELs) could enable  Mössbauer single-shot measurements without averaging, and a proof-of-principle demonstration has been reported.
However, so far, only a tiny fraction of all shots resulted in signal-photon numbers which are sufficiently high for a single-shot analysis.
Here, we demonstrate coherent nuclear-forward-scattering of self-seeded XFEL radiation, with up to 900 signal-photons per shot. We develop a sorting approach which allows us to include all data on a single-shot level, independent of the signal content of the individual shots. It utilizes the presence of different dynamics classes, i.e. different nuclear evolutions after each excitation. Each shot is assigned to one of the  classes, which can then be analyzed separately. Our approach determines the classes from the data without requiring theory modeling nor prior knowledge on the dynamics, making it also applicable to unknown phenomena. 
We envision that our approach opens up new grounds for M\"ossbauer science, enabling the study of out-of-equilibrium transient dynamics of the nuclei or their environment.
\end{abstract}

\maketitle

\iffalse
{\color{blue}Changes needed in the text for response:

\begin{itemize}
%\item general discussion of XFEL cases (high mean - sorting not required, average mean - current case, low mean - synchrotron) improve where we have it already

\item improve science case /applications in intro

\item In simulated data fig include 20, 200, 2000
   
\end{itemize}
}

\fi

%%%%%%%%%%% CONTENT %%%%%%%%%%%%%%%%%%%%%

\section{Introduction}

The exceptional resolution of Mössbauer spectroscopy results from the narrow linewidth of the nuclei and the recoilless interaction of photons in condensed matter due to the Mössbauer effect~\cite{mossbauer1958kernresonanzfluoreszenz}. It forms the basis for their widespread application relying on high energy resolution~\cite{moessbauer_story}. 
In the traditional setup, radioactive sources are used to record spectra over extended measurement times~\cite{chemistry}. Synchrotron radiation sources can considerably speed up the measurement and allow for measurements in the time domain~\cite{hf-review,Hastings1991,Roehlsberger2005},
but the narrow spectral width nevertheless implies the need for averaging over many x-ray shots. The reason is that the signal rate per shot is low ($\sim 10^{-3}$~ph/(pulse$\cdot \gamma_0$)) since the spectral width of the x-ray light is orders of magnitude broader than the nuclear resonance, such that the vast majority of the intense initial x-ray light is off-resonant and hence not contributing to the excitation.

This averaging requirement poses a severe challenge for  studying  statistical or non-repetitive dynamics, out-of-equilibrium dynamics or related phenomena with M\"ossbauer spectroscopy~\cite{10.1007/978-3-540-78697-9_16}.
If the sample dynamics is not always the same after each x-ray shot, then the averaging inevitably ranges over different evolution pathways from the out-of-equilibrium state back into equilibrium, thereby impeding their disentangling and understanding. 

For instance, statistical nuclear dynamics are expected  in the random sequence of coherent and incoherent emissions from a higher-excited nuclear ensemble, potentially involving the transient formation of entanglement~\cite{PhysRevA.59.1025,Moehring2007,PhysRevLett.99.193602}. More generally, M\"ossbauer nuclei are widely used to probe the dynamics of their surrounding host material~\cite{moessbauer_story_dreams,chemistry,hf-review}, with even non-repetitive host dynamics  being  mapped onto the nuclear dynamics. Particularly following external stimuli~\cite{PhysRevLett.77.3232,Heeg2021,PhysRevLett.82.3593,Vagizov2013,sakshath2017optical,doi:10.1126/sciadv.abc3991} the nuclear dynamics can change drastically. In the simplest case, the stimulus induces  dynamics only with a certain success probability. More interestingly, the host dynamics can involve  quantum mechanical superpositions, probabilistically leading to  different measurement outcomes.
Examples include the laser pumping of the electronic sub-system of the host~\cite{PhysRevLett.82.3593}, which can also be prepared in a superposition state.  Laser-pumping can also control the vibrational state of the host~\cite{Vagizov2013}, or the switching of spin-crossover complexes~\cite{sakshath2017optical,doi:10.1021/acs.jpclett.0c03733}. Analogously, magnons have been  studied via their transient impact on the nuclear dynamics~\cite{doi:10.1126/sciadv.abc3991}.  In chemistry~\cite{chemistry}, M\"ossbauer spectroscopy is pivotal, e.g., for studying biochemical reactions involving iron~\cite{VolkerSchunemann_2000}, but the study of short-lived transient intermediate states is difficult using M\"ossbauer spectroscopy~\cite{krebs2005rapid}.

Here, we demonstrate the disentangling of different M\"ossbauer  dynamics by analyzing  time-domain data on the single-shot level. For this, we tackle the averaging challenge from two sides.  First, we demonstrate  coherent nuclear forward scattering of self-seeded radiation delivered by the European XFEL~\cite{decking2020mhz,liu2023cascaded}, which allowed us to record M\"ossbauer datasets with up to 900 signal photons per x-ray shot using the isotope ${}^{57}$Fe. These shots with highest photon number directly reveal the nuclear dynamics  up to about 50~ns after excitation. Second, we develop an approach which generalizes the analysis to  shots with lower signal photon rate, and to longer times after excitation. It relies on the presence of distinct evolution pathways from out-of-equilibrium  back into equilibrium, which we denote as dynamics classes. We determine the classes from the data, and assign each shot to a class. Analyzing the shots of each class separately then avoids an averaging over different dynamics. The purely data-driven approach does not involve theoretical modeling, such that it applies to a priori unknown effects.

Increasing the signal photon rate until sufficient time-domain data can be recorded using single x-ray shots~\cite{10.1007/978-3-540-78697-9_16} is a  seemingly simple solution to the averaging challenge.
XFELs~\cite{emma2010first,barletta2010free,ishikawa2012compact,decking2020mhz} are routinely used for single-shot measurements involving electronic resonances~\cite{Bostedt2016,Rossbach2019,annurev:/content/journals/10.1146/annurev-physchem-032511-143720} and have recently moved this approach also within reach  for  M\"ossbauer resonances. As a first proof-of-principle experiment, Chumakov and co-workers extracted the nuclear hyperfine splitting in ${}^{57}$Fe from single-shot data with about 60 signal photons~\cite{chumakov2018superradiance}.
Meanwhile,  self-seeding enabled a further increase in the resonant x-ray flux~\cite{amann2012demonstration,inoue2019generation,liu2023cascaded,nam2021high}. This way, recently the ultra-narrow M\"ossbauer transition in ${}^{45}$Sc could resonantly be excited~\cite{shvyd2023resonant}. However, the corresponding observation of coherent nuclear forward scattering, which is the key requirement for most applications of M\"ossbauer nuclei, has not been reported before.

\begin{figure}[t]
    \centering
        \includegraphics[width=\columnwidth]{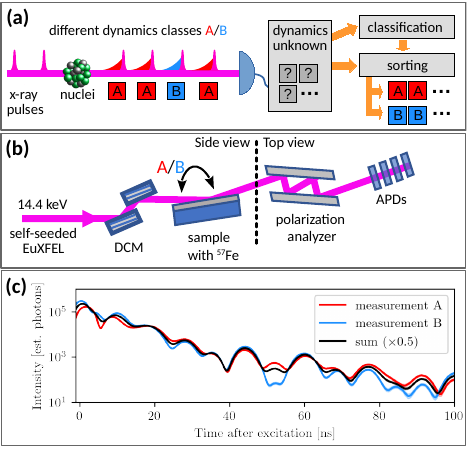}
        \caption{(a) Single-shot sorting. We consider a generic experiment  in which  M\"ossbauer nuclei may undergo different dynamics following an x-ray excitation. Two dynamics classes ``A'' and ''B'' are indicated in red and blue as examples. The information on the dynamics is typically lost for most shots during the measurement (gray squares). Our data-driven approach identifies different dynamics classes, and subsequently sorts all shots according to the identified classes. This way, the classes can be analyzed separately, avoiding an averaging over different dynamics.
        (b) Schematic experimental setup at European XFEL. The self-seeded x-rays pass through a double-crystal monochromator (DCM), which removes the off-resonant SASE background, and are then reflected from a thin-film waveguide containing ${}^{57}$Fe nuclei. The nuclear-resonant signal and the off-resonant background are separated using a polarization analyzer. The time-dependent intensity of the x-rays scattered by the nuclei is then recorded using avalanche photo diodes (APD). The two dynamics classes are deterministically implemented using slightly different scattering geometries, as explained in the main text.
        (c) Average intensities as a function of time after x-ray excitation for dynamics classes A (red) and B (blue) separately, as well as their average (black). Only the latter signal is accessible without per-shot information on the dynamics. The 1-sigma uncertainty band of the measurements falls mostly within the width of the plotted lines (see Appendix~\ref{detection}).
        }
        \label{fig:overview}
\end{figure}

% EDITED: moved extended data figure to main text because of new format
\begin{figure}[t]
    \centering
    % EDITED new figure with analysis ROI
    \includegraphics[width=0.8\linewidth]{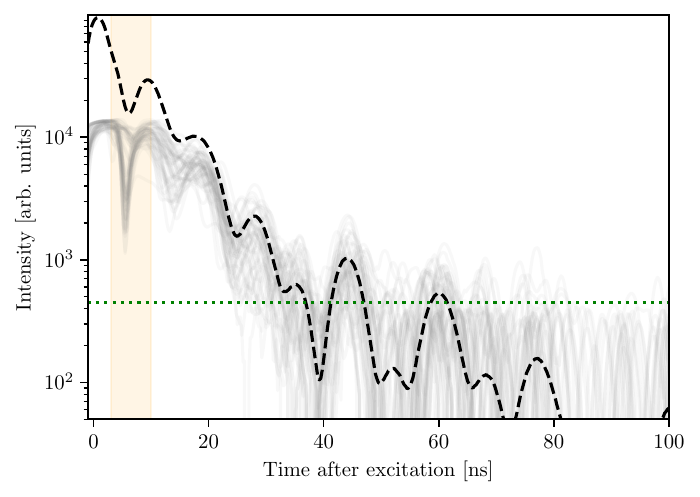}
    \caption{Example raw data with high signal content.
    Out of the full dataset, the time-dependent intensities of the {50} shots with highest signal content are shown in gray. The lines are plotted semi-transparent. This way, for example, the individual single-photon detection events at late times can be distinguished from the overlapping data at early times via the plot density.  The black dashed line displays the corresponding time-dependent intensity averaged over the full dataset.  The green dotted line indicates the average APD signal amplitude for individual recorded signal photons. The orange shaded area indicates the analysis ROI between 3\,ns and 10\,ns identified by the data-driven algorithm.}
    \label{fig:exampleHighNTraces}
\end{figure}

Despite this progress, recording M\"ossbauer single-shot datasets beyond the proof-of-principle stage remains a challenge, since due to the XFEL photon-number distribution \cite{gorobtsov2018seeded} the shots with highest signal-photon number only form a negligible fraction of all shots~\cite{chumakov2018superradiance}. The vast majority of shots typically does not have sufficient signal for a single-shot analysis. Increasing the x-ray intensity further is subject to sample damage, since the ``probe-before-destroy'' paradigm~\cite{Neutze2000} established for electronic scattering cannot straightforwardly be generalized to Mössbauer nuclei due to the long lifetimes of the nuclear excited states. This is particularly important  if focusing to smaller excitation volumes is desired in order to avoid extensive spatial averaging. Recording M\"ossbauer single-shot data is further limited by the dynamical range of the commonly-used avalanche photo-diode (APD) detectors, since the overall exponential decay of the scattered photon intensity typically spans over several orders of magnitude, thereby exceeding the linear detection range. As an example, in the present experiment, we observe APD saturation effects at early times in the shots with highest signal-photon number, even though they resolve the nuclear dynamics only for about 50\,ns after excitation. % beyond about one third of the natural lifetime of ${}^{57}$Fe  (see Methods Section ``Raw data'').
Another challenge is the desired scope of single-shot measurements, which is to probe non-repetitive phenomena. In~\cite{chumakov2018superradiance}, a known theory was fitted to the data in order to extract a single parameter,  the frequency of an oscillation throughout the entire time-domain. By contrast, lifting the averaging requirement is particularly important if the  phenomenon under study has a comparably subtle effect on the time-domain data. Further, the requirement of fitting theory models restricts the analysis to known effects.

%We overcome these challenges in directly recording and evaluating M\"ossbauer single-shot datasets using  a sorting approach on the single-shot level. 
Our experiment operates in an intermediate regime, where there exist a few outliers with highest photon number. The vast majority of shots has medium or low photon numbers, but nevertheless contributes significantly to the overall statistics (see Appendix~\ref{lowSignal}). In this intermediate regime, we overcome the above-mentioned challenges in directly recording and evaluating M\"ossbauer single-shot datasets using  a sorting approach on the single-shot level. 
Note that sorting approaches  are well-established, e.g., in the context of coherent diffractive imaging~\cite{Spence_2012,Schlichting:it5004,Dold2025,Zimmermann2023}, where  the scattering data is used to determine the sample orientation on a per-shot basis. However, this classification is concerned with geometric properties, and not with dynamics.
Another example is coherent correlation imaging~\cite{Klose2023},  which  classifies recorded camera images in Fourier space. 
Measure-and-sort approaches are also used involving auxiliary diagnostics measurements, e.g., to determine the unknown delay between optical pump  and XFEL probe pulses~\cite{harmand2013achieving}, or to characterize XFEL pulses temporally~\cite{hartmann2018attosecond,funke2024capturingnonlinearelectrondynamics} and spectrally~\cite{PhysRevLett.129.183204}. However, analogous approaches have not been explored with nuclear resonances before, and standard XFEL diagnostics does not straightforwardly apply to nuclear resonances because of their ultra-narrow spectral width, which typically cannot be resolved.

\begin{figure*}[t!]
    \centering
        \includegraphics[width=0.9\textwidth]{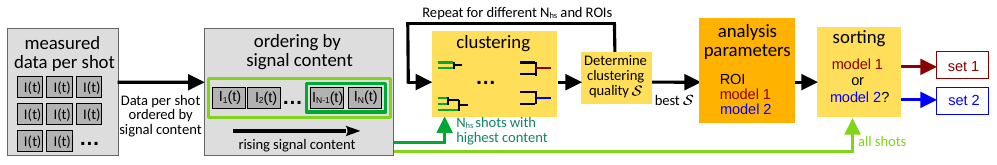}
        \caption{Overview of the clustering and sorting. First, all shots are sorted by their signal content. Second, the $N_\textrm{hs}$ shots with highest signal content are used} to identify different dynamics classes using a clustering algorithm. This is repeated for different numbers of high signal shots $N_\textrm{hs}$ and regions of interest (ROI) in which the analysis is performed. For each set of parameters, the clustering quality $\mathcal{S}$ is determined. From the parameter set that leads to the best clustering quality $\mathcal{S}$, two models are extracted as well as the optimal ROI. The resulting two models are then used as a reference to sort all shots of the experiment into corresponding sets which can then be analyzed separately.
        \label{fig:algorithm}
\end{figure*}

\section{Results}

{\bf Experimental setup.} The schematic setup of our approach is illustrated in Fig.~\ref{fig:overview}(a). We consider  a generic M\"ossbauer experiment at an XFEL source (Fig.~\ref{fig:overview}(a)) comprising a large number of shots, each initiated by the arrival of an x-ray pulse. After each x-ray excitation, detectors record the time-dependent scattered light intensity as a signature of the nuclear dynamics. As an example, one of two dynamics classes (A or B) is randomly realized in each shot. For most shots, the information on the dynamics class is not accessible by standard measurement approaches due to insufficient signal photon statistics, as indicated by the gray squares.
Our goal therefore is to sort the data on a single-shot level such that the dynamics classes can be distinguished and analyzed separately.

In order to demonstrate our approach, we performed an experiment at the Materials Imaging and Diagnostics (MID) instrument of the European X-ray Free Electron Laser~\cite{madsen2021materials}. The schematic setup is shown in  Fig.~\ref{fig:overview}(b). The XFEL was operated in self-seeding mode~\cite{liu2023cascaded} with an average pulse energy of $170\,\mu$J (about $35\%$ self-seeding fraction) and  bandwidth of 1.2\,eV. The accelerator was operated in the 2.2\,MHz mode, with 30 pulses per train and 440\,ns separation between two subsequent pulses, which matches the nuclear life time very well.  The photon energy was set to the nuclear resonance energy of 14.4 keV employing absolute energy measurements via the Bond method~\cite{Bond:a02913}. The incoming self-seeded x-rays pass MID's Si(111)  double-crystal monochromator (DCM) with about 1\,eV transmission bandwith, in order to remove the off-resonant SASE 
(self-amplified spontaneous emission) background. The x-rays are intrinsically polarized in the horizontal plane, and a channel-cut Si(840) polarization analyzer \cite{PhysRevLett.70.359,Marx-Glowna:hf5410,Roehlsberger2005}
in crossed setting is used to block this polarization channel, thereby protecting the avalanche photo-diode (APD) detectors  from the incident beam. The  ${}^{57}$Fe M\"ossbauer nuclei are embedded in a thin-film waveguide sample (see Appendix~\ref{sample}) aligned to reflect the beam in grazing incidence geometry. A weak external magnetic field is applied to align the nuclear magnetization along the beam propagation direction, so that the circular polarizations are eigen-polarizations and the incident horizontal linear $\sigma$-polarization experiences strong scattering into the perpedicular $\pi$-polarization. This  polarization component will pass the analyzer and its time-dependent intensity is recorded with APDs. 

{\bf Raw data.}
A detection signal averaged over 362.610 shots
 is shown in Fig.~\ref{fig:overview}~as the black line. Example single-shot APD raw data with highest signal content are shown in Fig.~\ref{fig:exampleHighNTraces}. The black-dashed line indicates the average over the entire dataset as a reference. While at early times, saturation effects are visible in the high-signal-content shots, at times later than approximately 50\,ns their statistics is at most on the single-photon level and a single-shot analysis is not possible anymore. Nonetheless, at around 5\,ns after excitation, the shots visually separate into two groups, whose time-dependent intensities differ more than their spread. This already indicates the presence of two different dynamics classes.  More details  are provided in Appendix~\ref{raw}.
 In addition, in Fig.~\ref{fig:exampleHighNTraces} example shots with high, medium and low signal content are displayed.

{\bf Experiment design.} To allow for a quantitative assessment of our approach, we designed the experiment such that it features two different dynamics classes, and that it is known which shot belongs to which class. To this end, two separate measurements are performed using a single target, but for two slightly different incident angles of the x-rays onto the waveguide, which represent the two dynamics classes (indicated as A/B in Fig.~\ref{fig:overview}(b); see Appendix~\ref{sample}). 
% EDITED: removed 
%This way, 
The time-domain data of each of the two classes can be obtained separately by averaging over the  data of each measurement for later reference. This information is blinded in the actual analysis. The two measured reference datasets averaged over all shots of each subset are shown in Fig.~\ref{fig:overview}(c). They can be well-described by standard low-excitation theory (see Appendix~\ref{sample}), and exhibit  characteristic differences, e.g., at around 5\,ns, 25\,ns, 50\,ns, and after 70\,ns. Note that the measured data are not histograms of signal photon arrival times, as they are established, e.g., for nuclear resonant scattering experiments at synchrotrons. Rather, they are sums of the full APD detector traces, since a disentangling of shots with tens or even hundreds of signal photons into individual arrival times is challenging (see Fig.~\ref{fig:exampleHighNTraces}).

\begin{figure}[t]
    \centering
    \includegraphics[width=\linewidth]{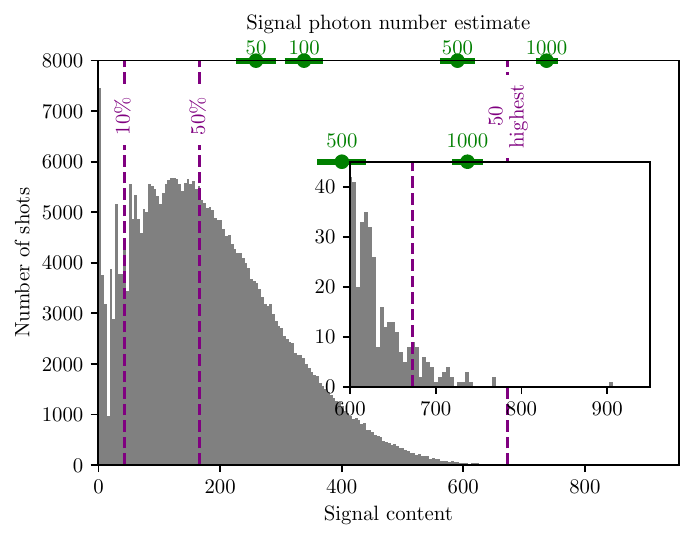}
    \caption{ Histogram of signal contents.
    A histogram of the signal content of all recorded individual shots. Estimates on the corresponding photon numbers and their uncertainties are indicated by the green bars on the top axis. For details on the photon number estimate see Appendix~\ref{content}. In addition, the purple dashed lines indicate low (10\% of the traces), medium (50\% of the traces) and high (50 highest traces) signal content and examples for those regimes are shown in Fig.~\ref{fig:exampleHighNTraces}. The inset shows a magnification of the highest signal-content region.}
    \label{fig:signalContent}
\end{figure}

{\bf Single-shot sorting approach.}
For the analysis, we do not use any of the  prior knowledge about the measurement geometry.
% EDITED removed text
Instead, the information on the dynamics class is eliminated by combining the  different shots of the two measurements into a single dataset, with its average shown as the black line in Fig.~\ref{fig:overview}(c). The analysis sequence is illustrated in Fig.~\ref{fig:algorithm}.
The first step is to identify dynamics classes (model 1 and 2) in the ``blinded'' data (indicated by the gray rectangles), which can then be used to sort all shots accordingly. For this clustering step, only a selection of the shots with the highest signal content is used, which circumvents the statistical uncertainty in the data associated with low-count shots. This way time-domain data representing the specific features of each class can be extracted more reliably from the measured data. In the following we denote this step as model-building. Generally, M\"ossbauer time-domain data features an approximately exponentially decaying scattered light intensity, due to the spontaneous emission of the nuclei. Shots suitable for the model-building should ideally contain information, i.e., recorded signal photons, across a broad time range after excitation, which is why the logarithm of the APD trace is applied to linearize this scaling. The signal content is then defined as sum over the logarithm of the APD trace (see Appendix~\ref{content}). To avoid prompt artifacts from the intense incident x-ray pulse, the first 3\,ns are excluded from the analysis (see Appendix~\ref{raw}). A histogram of the signal content of all shots in the combined dataset is displayed in Fig.~\ref{fig:signalContent}. 
The signal content fluctuates from shot to shot due to the natural photon number fluctuations and shots with highest signal content are extremely rare. In addition, the signal content has a non-linear correlation with the photon number.

{\bf Clustering step.}
In order to search for the presence of different dynamics classes, we employ an unsupervised machine-learning approach in the form of an agglomerative clustering algorithm~\cite{ward1963hierarchical} on the set of the $N_\textrm{hs}$ shots with highest signal content. The clustering is based on calculating the pairwise distances between the  $N_\textrm{hs}$ shots. 
As a distance measure $d(I_i,I_j)$  between the APD traces of two shots $I_i$ and $I_j$, we use the negative Poissonian log-likelihood $P(I_i,I_j)=-\int_\text{ROI}\text{d}t \,  I_i(t) \cdot \ln{I_j(t)} - I_j(t) - \ln{\Gamma(I_i(t)+1)}$ integrated over the analysis time ROI, symmetrized by using the maximum of both directions $d(I_i,I_j)=\text{max}\left[P(I_i,I_j), P(I_j,I_i)\right]$ 
and evaluated within an analysis region of interest (ROI) on the time axis. As illustrated in the ``clustering''  box in Fig.~\ref{fig:algorithm}, the two closest shots are joined into a single cluster. This clustering step is then repeated until only two clusters remain (For the result when enforcing three clusters see Supp. Fig.~\ref{fig:stabilityAnalysis} and Appendix~\ref{stability}). We employ the complete-linkage measure~\cite{johnson1967hierarchical} to quantify the distance between two clusters, which is defined by the maximum of the pairwise distances between individual elements of the two clusters. 

{\bf Clustering quality.} Next, we assess if the data indeed comprises different dynamics classes, by evaluating the quality and consistency of the clustering. For this, we use the silhouette score~\cite{rousseeuw1987silhouettes}, which quantifies how similar a shot is to the other elements of its own cluster, on a scale given by its distance to the other clusters. The measure compares the distance $\bar{d}_i$ of shot $i$ to the averaged time-domain data of its own cluster, with the respective distance $\bar{d'}_i$ to the second cluster. The silhouette score of a 
shot $i$ then is evaluated as $s_i = (\bar{d}_i - \bar{d'}_i)/\mathrm{max}\{\bar{d}_i, \bar{d'}_i\}$, and thus ranges from $-1$ to $+1$. A high score requires that the distance of shot $i$ to its own cluster is much smaller  than  the distances to the shots of the other clusters, $\bar{d}_i \gg \bar{d'}_i$.
An example of the silhoutte scores of the individual shots is displayed in Supplemental Fig.~\ref{fig:silhouette}.
A high mean silhouette score $\bar s_c$ averaged over all elements in cluster $c$ indicates that indeed a separate dynamics class was successfully identified in the data. As a measure for the  overall clustering quality $\mathcal{S}$ of both clusters, we use the minimum of the average scores of the two clusters $\mathcal{S} = \mathrm{min}_{c\in\textrm{clusters}}\{\bar s_c\}$ 
to ensure that both dynamics classes are well-represented. This clustering quality measure generalizes to more clusters and can be used to identify the optimal number of clusters, which in our case results in two clusters as expected.

\begin{figure}
    \centering
    \includegraphics[width=0.8\linewidth]{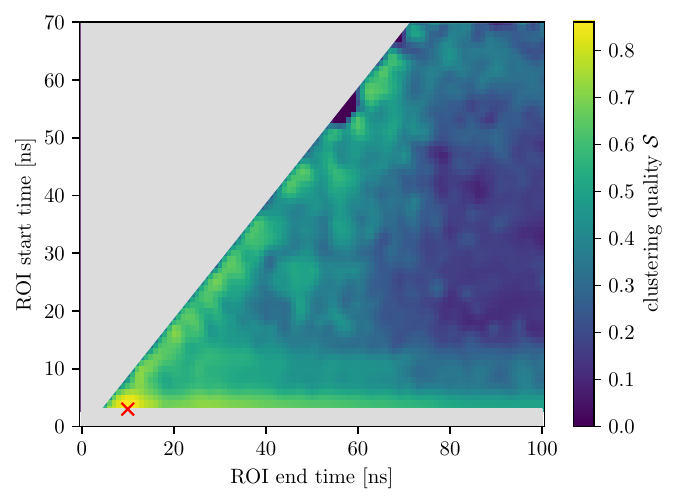}
    \caption{Overall clustering quality $\mathcal S$ as function of the analysis region of interest start and end times. Results are shown for the subset of $N_\textrm{hs}=50$ shots with highest signal content. A moving Gaussian average filter of width $\sigma=1$ns was applied to the data to reduce the influence of outliers and to favor stable analysis parameter regions. The lighter the color, the better the clustering quality $\mathcal{S}$, with the optimum found for the analysis ROI of (3\,ns, 10\,ns) and marked by the red cross. In gray invalid regions are marked.}
    \label{fig:trianglePlot}
\end{figure}

To determine the optimum analysis parameters, i.e., the number of highest-signal shots $N_\textrm{hs}$ and the analysis ROI, we maximize the clustering quality $\mathcal{S}$. For this, the above analysis is systematically repeated for different numbers of shots $N_\textrm{hs}$ and analysis ROIs, which takes approximately 6.5 CPU-hours on DESY's Compute-Cluster ``Maxwell''.
This way, the analysis parameters for the model building are fixed based on the experimental data without the need for  prior knowledge about the system, nor theory modeling of it. As an example, we show the overall clustering quality $\mathcal{S}$ as function of start- and end time of the analysis ROI for $N_\textrm{hs}=50$ shots in Fig.~\ref{fig:trianglePlot}. In order to reduce the influence of outliers and to favor stable analysis parameter regions, a narrow moving Gaussian average filter of width $\sigma=1$ns is applied to the clustering quality $\mathcal{S}$ in this figure. We find that best clustering qualities are grouped in certain ROI ranges, and we select the optimum ROI via the maximum of the clustering quality,  indicated by the red cross in Fig.~\ref{fig:trianglePlot}. Early times for the ROI are likely favored since the signal rate rapidly drops with time after excitation due to the exponential decay. As the result of the algorithm, we found that the clustering quality $\mathcal{S}$ is maximized by including the $N_\textrm{hs}=50$ shots with highest signal content out of all 362,610 shots into the model building, and by using an analysis ROI between 3~ns and 10~ns. Note that the choice of the optimal $N_\textrm{hs}$ is influenced by the statistic nature of the highest signal counts, and the clustering quality and the optimum ROIs are similar over a broader range of  $N_\textrm{hs}$, which indicates that the analysis is stable against variations in the analysis parameters.

After having identified the optimum analysis parameters based on the clustering quality $\mathcal{S}$, a further consistency check is possible by  analyzing the individual silhouette scores $s_i$ of the different shots in all clusters. Indeed, we find that our approach identified two  clusters  comparable in size, and that the majority of shots in both clusters has a high individual silhouette score (see Appendix~\ref{score} and Supplemental Fig.~\ref{fig:silhouette}). 

In general, the success of the clustering depends on the availability of high signal shots, as well as on the shape and the temporal position of the feature. In the Appendix~\ref{criterion}, we derive a criterion for the  signal photon number required for distinguishing dynamics classes based on a generic temporal feature. As expected, the required photon number depends on the start time of the temporal feature after the x-ray excitation, its duration, and its amplitude. We also verified the predictive power of the criterion based on the experimental data for the temporal feature analyzed here. Using numerically simulated data with higher photon numbers 
(see Appendix~\ref{simulated}), we also verified the criterion for temporal features at later times after excitation (see Supplemental Figs. 5 and 6). revision

{\bf Sorting step.} Once the dynamics classes are identified and corresponding models are generated, we proceed with the entire set of measured shots, independent of their signal content, and assign each shot to one of the models (see ``sorting'' in Fig.~\ref{fig:algorithm}). 
% EDITED: Removed 
%Note that 
This comparison of individual shots with the generated models has a crucial advantage over the direct comparison of individual shots. A direct comparison of two shots with low signal content is subject to comparably high statistical uncertainties. In contrast, the statistical fluctuations are greatly reduced in the models comprising a number of shots, such that a comparison of a low-signal shot with one of the models is more reliable. For this comparison, we employ the negative Poissonian log-likelihood $P(a,b)$, regarding the individual shots as the data to be tested against the models. This sorting step results in  one set of shots for each of the dynamics classes determined throughout the sorting.

\begin{figure}[t]
    \centering
    \includegraphics[width=\linewidth]{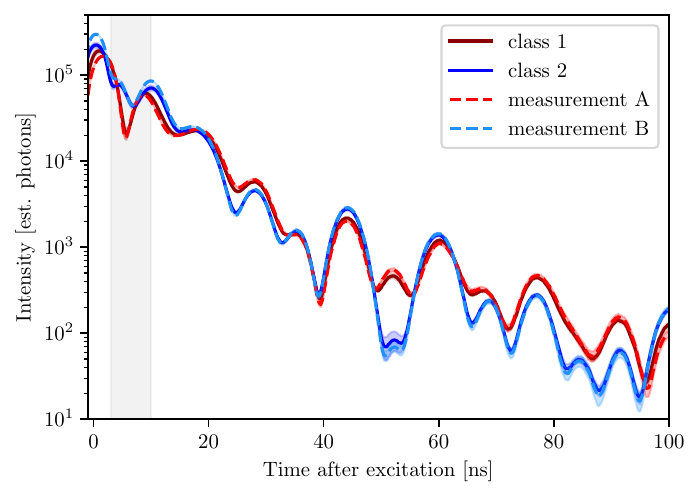}
    \caption{Results of the single-shot sorting analysis. 
    The results of the data-driven single-shot analysis (class 1 and 2) are shown as solid lines. The uncertainties are drawn as the shaded band around the lines. They are estimated based on the standard deviation of the subset analysis in Appendix~\ref{stability}.
    The dashed lines show the time-domain data measured in the experimental settings \rA (red dashed) and  \rB (blue dashed), which serve as a reference to benchmark the success of the sorting analysis. The $1\sigma$ uncertainty band of the measurements is indicated by the shaded area around the respective lines (see Appendix~\ref{detection}).     
    The two time-domain datasets are scaled for better comparability by fitting an overall scaling factor from 0ns to 100ns. The ROI (3-10\,ns) used for the analysis is indicated in gray. } 
    \label{fig:spectra-datadriven}
\end{figure}

The two sets of sorted data can now be analyzed individually. Summing over each set separately results in time-dependent intensities, which reflect the nuclear dynamics in the two respective dynamics classes, and form the main results of the analysis. They are shown as solid lines in Fig.~\ref{fig:spectra-datadriven}. 
The two time-dependent intensities clearly differ at certain times, e.g., around 5\,ns and 50\,ns as well as from 70\,ns to 90\,ns, indicating that the clustering algorithm identified different dynamics classes.

{\bf Verification of the single-shot sorting.}
In order to verify the data-driven analysis approach, we finally unblind the information on the experimental settings for each of the shots, which we have due to the specific design of the underlying experiment, but which was not used  during the data-driven analysis. The resulting ``true'' reference time-domain data for the two different dynamics classes  are shown as dashed lines in Fig.~\ref{fig:spectra-datadriven}.
We find that the results of the data-driven analysis indeed follow  the reference measurements closely. In particular, the  regions with characteristic higher deviations  between the two dynamics classes are well-reproduced in both cases. Interestingly, the regions around 50ns, or after 70ns, are well-recovered even though they are far outside the analysis ROI from 3ns to 10ns used to generate the original models.
We attribute deviations at initial  times up to 4\,ns to a preferred assignment of shots with lowest signal content to class 1, due to its lower integrated intensity in the ROI. This results in an over- (under-)compensation of saturation effects in class 1 (2) at early times, see Appendix~\ref{raw}.

 \begin{figure}
    \centering
    \includegraphics[width=\linewidth]{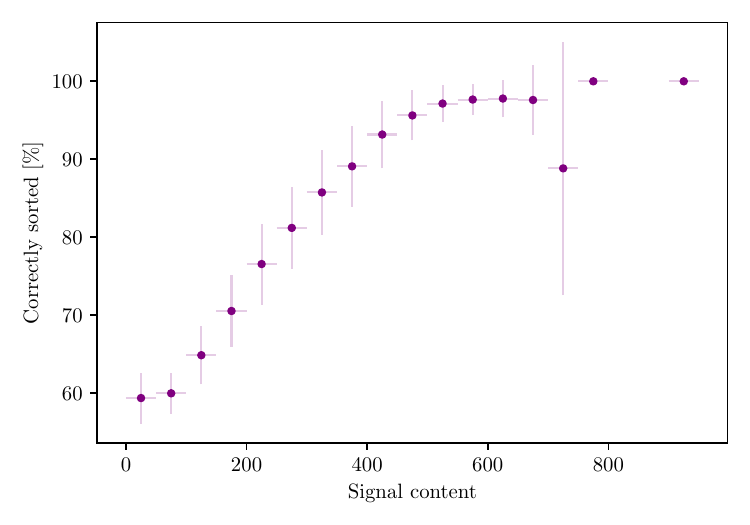}
    \caption{
    { Fraction of correctly sorted shots}. 
    The percentage of correctly sorted shots as function of the signal content. The horizontal error bars represent the widths of the corresponding signal content bins, while the vertical error bars indicate the $1\sigma$ uncertainty after averaging over all 50 subsets introduced in the Appendix~\ref{stability}. For the average the subsets are weighted by the respective number of shots in the bin. The vertical error grows towards highest signal content since the corresponding number of shots within a given subset progressively reduces, see Fig.~\ref{fig:signalContent}, until all shots are sorted correctly.
    }
    \label{fig:QualityOfSorting}
\end{figure}

Our experiment design also allows us to quantify how many of the shots were sorted into the correct dynamics class, since the latter is known for each shot. In Fig.~\ref{fig:QualityOfSorting}, the sorting accuracy is displayed as a function of the signal content (for the uncertainties, see the Appendix~\ref{stability}).
As expected, towards vanishing signal content, the fraction of correctly sorted shots approaches 50\%, corresponding to a random assignment of the shots to either of the two classes. Towards higher signal content, the ratio approaches 100\%, i.e., essentially all shots with such high signal content are correctly sorted into their respective class.
Furthermore, 1/3 of the total signal photons were recorded in shots with medium signal content between 200 and 500 (see Appendix~\ref{lowSignal}). These are too weak for the model-building, but on average are sorted correctly into their respective classes with a high probability. This shows that  the weaker shots indeed contribute significantly to the total data statistics in our sorting approach.

We also confirmed  that the single-shot sorting approach generalizes to different temporal features at later times, using numerically simulated data obtained by sampling the measured time-dependent intensities with the experimentally observed photon number statistics, but for a  higher mean photon number selected according to the single-shot resolution criterion (see Appendix~\ref{simulated}). 

\section{Discussion}

We have demonstrated the possibility to perform coherent nuclear forward scattering experiments using self-seeded XFEL radiation. The resonant peak flux (per x-ray pulse) achieved in the present experiment exceeds typical values accessible at synchrotron radiation facilities by about four orders of magnitude (See Appendix~\ref{excitation}). Interestingly, the high repetition rate at EuXFEL also renders the average resonant flux (per second) comparable to synchrotron conditions, and further improvements are within the present capabilities of the facility~\cite{liu2025updates}.

The exceptionally high resonant flux per shot unlocks fundamentally new applications for M\"ossbauer nuclei. In particular, our results demonstrate the possibility to disentangle different evolution pathways from the out-of-equilibrium state back into equilibrium by analyzing time-domain datasets on the level of single shots. For this, we  identify different dynamics classes and sort the data accordingly. Since the clustering analysis is data-driven and does not require a theoretical modeling of the data, also previously unknown phenomena can be searched for. The analysis can readily be applied to standard experimental approaches, without the need for dedicated measurements or instrumentation, also for more general signals beyond M\"ossbauer spectroscopy.
%
%Rather than analyzing each individual shot separately, our approach is to extract dynamics models from the data of the shots with the highest signal content. Subsequently, all shots can be sorted, i.e., assigned to one of the models. This way, also shots with lower signal-photon number can  contribute  to the analysis---even if their signal content is not sufficient for an individual analysis. 

Importantly, we found that our sorting analysis is capable of clearly revealing the differences in the two dynamics classes also for times larger than approximately 50\,ns after excitation. By contrast, even the rare shots with highest signal-photon number do not contain sufficient counts to resolve these differences in the dynamics directly, due to the overall approximately exponential decay of the scattered light intensity with time (see Appendix~\ref{raw}). This highlights the advantage of the sorting approach over the direct single-shot analysis.

By repeating the clustering analysis for different final numbers of clusters, the most probable  number of dynamics classes can be identified by the clustering quality measure $\mathcal{S}$.   In the present experiment, we find that an analysis with two clusters yields the highest clustering quality and stability, consistent with the original experimental design  (see Appendix~\ref{sample}). %We further applied our analysis to the data of only one of the measurement settings, which resulted in the prediction of a single dynamics class, as expected. 
In practice, already a stable and reproducible identification of qualitatively different time-domain data is  a clear indication that the experiment comprises different classes of dynamics  
(see Appendix~\ref{stability}).

We envision that the single-shot sorting approach applied to coherent forward scattering data recorded with self-seeded x-ray pulses will open up new applications for M\"ossbauer science. Broadly speaking, different dynamics classes can originate from the nuclei themselves, or from the surrounding host environment.
Under XFEL excitation conditions, even the unperturbed nuclear relaxation dynamics of a multiply-excited nuclear ensemble back to the ground state comprises different sequences of coherent and incoherent emission events. The time-resolved observation of the emitted photons can project the nuclear ensemble into entangled states~\cite{PhysRevLett.99.193602}, or correlate independent excitation volumes in the sample~\cite{PhysRevA.59.1025,Moehring2007}. An analysis on the single-shot level  provides a route towards the observation of the nuclear ensemble in a quantum state. Furthermore,  a major contribution to the incoherent decay is formed by internal conversion~\cite{doi:https://doi.org/10.1002/9781119348450}, such that our method provides an opportunity to study the interaction between the nuclei and the electronic environment in a controlled way.
On the other hand, the majority of current-day applications of M\"ossbauer nuclei are related to studying the structure and dynamics of the surrounding host material. The host evolution  may also impose different dynamics onto the nuclei, e.g., if it involves quantum-mechanical superposition states with probabilistic measurement outcomes. Our sorting approach opens an avenue to developing the nuclei  as non-invasive high-resolution probes for the electronic, phononic and structural dynamics of the environment geared towards  the unique excitation conditions at XFELs. At these machines, 
mostly interactions of the x-rays with electronic degrees of freedom are used to probe targets, while these electronic properties oftentimes are subject of the investigation themselves, such that the probe and the probed system are intrinsically intertwined. By contrast, nuclei form an independent subsystem,  and thereby offer valuable complementary information. Key examples involve the dynamics after impulsive pumping of the electronic~\cite{PhysRevLett.82.3593}, vibrational~\cite{Vagizov2013}, spin~\cite{sakshath2017optical,doi:10.1021/acs.jpclett.0c03733} or magnetic~\cite{doi:10.1126/sciadv.abc3991} state  of the host material. We further envision the extension of the sorting approach to nuclear inelastic scattering~\cite{Chumakov1998,Vagizov2013}, which next to the study of phonons could also allow one to generalize XFEL M\"ossbauer studies to liquids, gases, plasmas, higher-energy nuclear transitions, ``detect-before-destruct'' approaches, or other samples with low probability for recoilless x-ray-nucleus interactions~\cite{Baron_1996,PhysRevB.73.024203}. Furthermore, coherent control mechanisms with Mössbauer nuclei \cite{Heeg2021,doi:10.1126/sciadv.abc3991} could be transferred towards the single-shot domain.

\appendix

\section{\label{detection}Detection}

Since the individual shots contain up to several hundred signal photons, the majority of which arrives in a comparably short time interval after the x-ray excitation, the employed APDs cannot distinguish the individual arriving photons. Therefore, rather than counting arrival times, we record the entire time-dependent voltage traces of the APDs using fast digitizers. As a result, the recorded time-domain data is convoluted by the detector response. For isolated signal photons, the detector response can be modeled using a Gaussian envelope of about 2.5\,ns temporal full-width at half maximum. 
% EDITED: removed
%In order 
%
To estimate the uncertainty of the measurements, the recorded voltage traces are scaled with the area of the single-photon detector response, such that the area under the curve is equal to the estimated integrated number of photons. Subsequently, a Poissonian uncertainty can be applied.

Besides the APD traces, further auxiliary and diagnostics observables related to the experiment, especially the XFEL beam, are saved leading to a total data size of approximately 200\,GB for the two measurements together.

\section{\label{excitation}Excitation conditions}
In the present experiment, the 300 x-ray pulses per second with mean  energy of $170 \mu$J ($35\%$ self-seeding fraction with a spectral bandwidth of 1.2~eV) translate into average and peak spectral fluxes of $3\times 10^4$ photons/$(s\cdot \gamma_0)$ and 101 photons/$(\mathrm{pulse}\cdot \gamma_0)$. Here, $\gamma_0 = 4.7$~neV is the natural linewidth of the ${}^{57}$Fe M\"ossbauer transition at 14.4~keV. The recorded time spectra exhibit a superradiant line broadening to  $\Gamma\approx 10\,\gamma_0$, estimated from the decay of the time-dependent intensity by 3 orders of magnitude over approximately 80\,ns. 

For comparison, the XFEL experiment by Chumakov \textit{et al.}~\cite{chumakov2018superradiance} performed at SACLA without self-seeding reported $6.7\times 10^{10}$ photons/pulse in a 75 eV bandwidth, with a repetition rate of 30 pulses/second. This corresponds to   $126$ photons/(s$\cdot\gamma_0)$ and $4$ photons/(pulse$\cdot \gamma_0$). Regarding synchrotrons, for example, after the high-resolution monochromator at the Nuclear Resonance Beamline ID14 at the European Synchrotron Radiation Facility, $1.4\times 10^{10}$ photons/second are delivered in a bandwidth of $6.4$~meV at 14.4~keV, with a x-ray pulse spacing of 176~ns. This translates into $10^4$ photons/$(s\cdot \gamma_0)$ and $2\times 10^{-3}$ photons/$(\mathrm{pulse}\cdot \gamma_0)$. 

In comparison to state-of-the-art synchrotron conditions, the high-repetition self-seeding mode of EuXFEL allowed us to achieve a more than $10^4$ times higher number of resonant photons per pulse without loss in average number of resonant photons,  which forms the basis for the reported single-shot sorting analysis.

\section{\label{raw}Raw data} 

The digitized APD traces represent the measured resonantly scattered intensity as function of time (see Appendix~\ref{detection}). 
In Fig.~\ref{fig:exampleHighNTraces}, those APD traces for the 50 shots with highest signal content (see Appendix~\ref{content}) are displayed. For comparison, the reference time-domain data obtained by averaging over all shots is displayed as the black dashed line. There are three important observations. First, up to 20\,ns there is a strong deviation in the recorded intensity of the single shots as compared to the averaged time-domain data. This can be attributed to saturation effects in the detector, due to the limited dynamical range of the APDs employed in the experiment.  This saturation is most pronounced at early times due to the overall approximately exponential decay of the scattered intensity with time. Furthermore, the saturation is strongest for the shots with highest signal content shown in the figure. By contrast, the majority of shots has a lower signal content (see Fig.~\ref{fig:signalContent}), and therefore is less affected by the detector saturation. As a result, the averaged data deviates from the shots with highest signal content, and better represents the true amount of light scattered by the nuclei. 
Second, we can analyze whether the individual shots with highest signal content contain sufficient count statistics to allow for an analysis of the time-domain data on a single-shot basis. We find that this is indeed the case at short times, although bounded by the saturation effects discussed above. 
At times later than approximately  50\,ns after the excitation, the recorded single-shot intensities approach the level of individual recorded photons, such that the dynamics can no longer be reconstructed reliably from the individual time spectra alone. 
Finally, we find that at times around 5\,ns, the data of the subset of shots with highest signal content clearly divides into two different subsets. One exhibits a pronounced dip in the intensity, while the other subset remains approximately constant. This feature is reflected by the region of interest from 3\,ns to 10\,ns identified by the clustering approach presented in the main text. Other features distinguishing the two dynamics classes A and B in our measurement are visible at later times in the averaged time-domain data, but they cannot be resolved on the single-shot level due to the low statistics.
Around $t=0$~ns, the averaged time-domain data comprise prompt scattering contributions, which can be attributed to the finite polarization extinction of the analyzer, the convolution of the signal with the single-photon detector response, and to background photons. Instead, the distinct feature at around 5~ns also appears in the corresponding idealized theory calculation, and can be attributed to the first quantum beat due to the interference of different scattering channels between the involved hyperfine states.
 
\section{\label{content}Signal content} 
Because the signal content should not be dominated by the high statistics at early times, but should account for a signal spread out over the entire analysis region, the signal content is defined by the area under the logarithm of the APD trace $\int_{3\text{ns}}^{400\text{ns}} \ln I(t)\text{d}t$. A histogram of the distribution of signal content is displayed in Fig.~\ref{fig:signalContent}. 
In addition, in Fig.~\ref{fig:signalContent} an estimate on the photon number corresponding to a certain range of signal content is given. To this end, 1000 single-shot traces with a specific photon number $N$ are simulated by drawing $N$ random numbers from the averaged time-domain data and summing up the single-photon detector response of those. For those artificial traces the signal content is calculated. To avoid influence of the saturation effects, the signal content is evaluated only after 20\,ns. For each $N$ this results in an approximately Gaussian distribution. Finally, for all shots in our measured dataset, whose signal content from 20\,ns onwards lies within the 1$\sigma$ interval around the mean of the Gaussian distribution, the mean and standard deviation of the signal content over the whole time range is calculated. This is the estimated signal content and its uncertainty for a given photon number $N$, which is indicated by the green dots and errorbars in the figure.
It is important to note that the shots with highest signal content are statistical outliers, and the 50 shots shown above only constitute a tiny fraction of the 362.610 shots measured during the experiment (181.350 shots for A, and 181.260 for B). 
To illustrate this further, example shots with high, medium and low signal content are shown in Fig.~\ref{fig:exampleHighNTraces}. For the high signal content shots, the ten shots with highest signal content are used, for medium signal content the ten with highest signal content out of the lower half and for low signal content the ten highest out of the lowest 10\%. The respective cutoffs are shown in Fig.~\ref{fig:signalContent} as the dashed purple lines. In the high-signal single-shot data, the difference at around 5\,ns is clearly visible while later features already lack statistics. In the shots with medium signal content, the differences at around 5\,ns are still partially visible, while the low signal shots contain only a small number of photons.

\section{\label{sample}Sample and scattering geometries}  
%Si/Pt(14.82 nm)/${}^{57}$Fe(15.32 nm)/Pt(2.10 nm)
The measurements were performed on a thin-film cavity fabricated by sputter deposition on a silicon substrate. The layer structure from bottom to top was determined as Si/Pt(14.8\,nm)/${}^{57}$Fe(15.3\,nm)/Pt(2.1\,nm), via fits to  an electronic $\theta-2\theta$ reflectivity curve measured at the P01 High Resolution Dynamics Beamline at the synchrotron radiation source PETRA III (at DESY in Hamburg) using the software packages GenX~\cite{Glavic:ge5118} and Nexus~\cite{https://doi.org/10.5281/zenodo.13832946}. Using this structure, the incidence angles of the scattering geometry were then determined via fits to the time-dependent intensities measured at European XFEL using Nexus~\cite{https://doi.org/10.5281/zenodo.13832946}. In these fits, the finite temporal width of the APD photon detection signals is modeled using a convolution with a Gaussian detector response. The theory fits together with the experimental data are shown in Supp. Fig.~\ref{fig:rawDataFit}. The incidence angles are obtained as
$4.15$ mrad % 0.237895 deg
and 
$4.03$ mrad %0.2307759
in the two measurements, respectively. 
These are close to the first minimum in the electronic reflection curve, corresponding to the resonant driving of the first cavity mode. Additionally, the two datasets feature a slightly different canting angle of the analyzer crystal, due to a realignment between the two measurements.

\section{\label{score}Silhouette score} 
The assessment of the clustering algorithm used in the main text is based on the overall clustering quality $\mathcal{S}$. As a more detailed consistency check for the clustering, the individual silhouette scores of the shots used for the model building are shown in Supplemental Fig.~\ref{fig:silhouette} for the optimum analysis parameters $N_\textrm{hs} = 50$ and analysis ROI between 3\,ns and 10\,ns. Each horizontal bar represents the silhouette score of a single shot with the colors indicating the cluster, to which the shot belongs. For better visibility, the clusters are spatially separated. A large fraction of the shots has a silhouette score larger than 0.5, many even larger than 0.8, indicating good clustering \cite{kaufman2009finding}. 
A few shots have a small silhouette score indicating that they do not so clearly belong to cluster. Nonetheless, those outliers are only few and together with the fact that a stable ROI is found, which is similar for different number of high-signal content shots $N_\textrm{hs}$, the model generation is reliable.

\section{\label{stability}Stability of analysis} 
The stability, reproducibility and consistency of the analysis can be tested in various ways. First, one may analyze the two time-domain datasets reconstructed from the data in the main text. To this end, we artificially create repetitions of the experiment with different statistical realizations of the photon detection by dividing the full dataset randomly into 5 subsets. Note that this division at the same time corresponds to a reduction of the effective measurement time realized in the experiment by a factor of 5, thereby rendering the analysis more challenging. Subsequently, the complete sorting algorithm described in the main part is applied to each of the subsets. To improve the statistics of this analysis, we repeat this procedure  ten times with new randomly chosen subsets. 
Afterwards, the 50 reconstructed time-domain datasets can be compared to each other and their standard deviation at any instance in time  provides an estimate of the uncertainty of the analysis due to data variability. For the sorting analysis discussed in the main part, the uncertainties are indicated as the shaded areas around the respective lines in Fig.~\ref{fig:spectra-datadriven}. It can be seen that the uncertainties are small compared to the difference between the averaged time-domain data of the two measurement sets, such that they can reliably be distinguished.

Next, we can use a stability analysis to verify the identification of dynamics classes, as shown in Supp. Fig.~\ref{fig:stabilityAnalysis}. As a first test, we perform the analysis only on the shots belonging to one of the measurements, A or B, while still enforcing the determination of two clusters in the clustering analysis. The results are shown in the left panel (a) of Supp. Fig.~\ref{fig:stabilityAnalysis}. The red and blue solid lines display the recovered averaged time-domain data of the two classes. The dashed lines show the recovered time-domain data of Fig.~\ref{fig:spectra-datadriven} corresponding to class 1 (red dashed) and class 2 (blue dashed) as a reference. For better clarity, the two cases with data of measurement A and measurement B are shifted with respect to each other along the y axis. We find that the two recovered datasets  (test class 1 and test class 2) agree within their uncertainty, such that we conclude that the algorithm did not identify different dynamics classes. Furthermore, the recovered averaged time-domain datasets agree with the reference data  for measurement A in case that the shots of A were used for the analysis, and analogously for B. Thus, the analysis works as expected.
 
In a third test, we repeat the analysis in the main text with data of both measurements A and B, but allow for three clusters in the clustering analysis. The corresponding results are shown in the right panel (b) of  Supp. Fig.~\ref{fig:stabilityAnalysis}. As a first observation, we find that the overall uncertainties are considerably larger than in the case with two clusters (see Fig.~\ref{fig:spectra-datadriven}), indicating a less reliable analysis. Second, we find that two of the recovered time-domain data (test class 1 and test class 3) agree well with each other, within their respective uncertainties. They also agree well with the reference data ``class 1'' obtained with the analysis involving only two clusters in Fig.~\ref{fig:spectra-datadriven}. Correspondingly, the third of the recovered time-domain data (test class 2) agrees well  with the second reference data ``class 2''. Thus, the  analysis is consistent with only two dynamics classes in the data, as expected. 

Overall, we therefore conclude that the sorting algorithm works as expected, in particular with regards to the correct number of dynamics classes extracted from the data.

\section{\label{lowSignal}Impact of shots with lower signal content}
The $N_\textrm{hs}=50$ shots with highest signal content represent approximately $10^{-2}\,\%$ of all recorded shots. The strongest 10\,\% of the shots contain only approximately 1/3 of the total number of measured signal photons. This raises the question, how the remaining signal photons contribute to the analysis. 
Due to the specific experiment design, the true class label is known for each shot. 
Therefore, we can analyze the accuracy of the sorting algorithm, and thus the impact of individual shots to the overall data analysis statistics. In Fig. \ref{fig:QualityOfSorting}, the fraction of correctly sorted shots is shown 
as function of their signal content.  To estimate the significance of the shots with lower signal content, we consider shots which on average are correctly sorted with at least 2/3 probability (signal content higher than 200). As an upper bound, we choose a signal content of 500, beyond which the shots are almost always correctly sorted. We find that these shots with intermediate signal content contain approximately 1/3 of the total recorded signal photons.  Hence, already this rather conservative estimate confirms that the weaker shots contribute significantly to the total data statistics in our sorting approach, especially at early times thereby counteracting the saturation effects in the high-signal shots.

\section{\label{criterion}Single-shot resolution criterion}
The single-shot sorting approach presented in the main text readily generalizes to other settings. It relies on the identification of different dynamics classes. These  become distinguishable if their respective high-signal-content shots are  ``sufficiently different''. For time-dependent intensities as they are measured in our experiment, two classes can be distinguished if there is a time interval after excitation in which the signals of the two classes differ more than their respective uncertainties. This condition can be formalized into a criterion for the signal photon number necessary to distinguish two classes based on a specific temporal feature. We assume that the two classes differ in the time interval $\Delta t = t_E - t_S$ by a rectangular temporal feature with relative amplitude $A$. Further, we consider the overall exponential decay of the two classes with lifetime $\tau$, as it strongly impacts the required photon number. Hence, the average time-dependent count rates for the two classes are given by $I_1(t) = (N_\textrm{ph}/\tau)\,\exp(-t/\tau)$ and $I_2(t) = I_1(t)\,\left[1- A\,\Theta(t-t_S)\Theta(t_E-t) \right]$, where $N$ is the total number of signal photons. 
As a criterion whether the two classes can be distinguished, we demand that the 
integrated difference $D$ between the time-dependent intensities
\begin{align}
    D =& \left| \int_{t_S}^{t_E}I_1(t)dt - \int_{t_S}^{t_E}I_2(t)dt \right| %\\ % = \int_{t_1}^{t_2} AN \frac{exp(-t/\tau)}{\tau}dt 
%     =& A {N\left[\exp(t_1/\tau) - \exp(t_2/\tau)  \right]} \,,
\end{align}
exceeds its uncertainty $\Delta D$ 
\begin{align}
    \Delta D =& \sqrt{\Delta\left(\int_{t_S}^{t_E}I_1(t)dt\right)^2 + \Delta\left(\int_{t_S}^{t_E}I_2(t)dt\right)^2}% \\
    %=& \sqrt{\int_{t_1}^{t_2}I_1(t)dt + \int_{t_1}^{t_2}I_2(t)dt}\\
    %=& \sqrt{\int_{t_1}^{t_2}N(2-A)\frac{exp(-t/\tau)}{\tau}dt} \\
%    =& \sqrt{2-A} \sqrt{N\left[\exp(\frac{t_1}{\tau}) - \exp(\frac{t_2}{\tau})  \right]}  \, .
\end{align}
by a factor of 3 ($3\sigma$ interval). This leads to an estimate for the required signal photon number given by 
\begin{equation} \label{eq:singleShot_N}
    N_\textrm{ph}\geq \frac{9 (2-A)}{A^2} \frac{1}{\exp\left(-t_1/\tau\right) - \exp\left(-t_2/\tau\right)} \, .
\end{equation}
In the limit of a short temporal feature $\Delta t \ll \tau$, this expression can be approximated by
\begin{equation}
    N_\textrm{ph}\geq \frac{9 (2-A)}{A^2}\:\frac{\tau}{\Delta t} \:e^{t_1/\tau} \, .
\end{equation}
The single-shot resolution criterion applied to the experimental feature at around 5\,ns with $t_1=4$, $t_2=6$, $A=0.67$ and $\tau\approx 14$\,ns due to superradiant broadening results in $N_\textrm{ph}\geq 267$, which is in agreement with the experimental statistics. Similarly for the features around 25\,ns (50\,ns) we find $N_\textrm{ph}\geq 2097$ ($N_\textrm{ph}\geq 2415$) for $t_1=23\,$ns, $t_2=25$\,ns and $A=0.5$ ($t_1=50\,$ns, $t_2=54$\,ns, $A=0.8$). The statistics in the experiment are too small to resolve the latter two features in the clustering step as can also be seen from Fig.~\ref{fig:exampleHighNTraces}.

In order to explore the predictive power of the single-shot resolution criterion  Eq.~(\ref{eq:singleShot_N}), and to show that the single-shot sorting also applies to more general temporal features, we numerically simulate data. We use the measured time-dependent intensities as probability distributions, and draw the photon arrival times for shots with a desired signal photon number $\tilde{N}_\textrm{ph}$ from it. We then convolve the arrival-time histogram with the measured single-photon APD response to obtain a simulated shot. We repeat this step 50 times for each of the two classes, resulting in 100 simulated APD traces. Afterwards, we apply the clustering algorithm described in the main text to this data, and evaluate the clustering quality $\mathcal{S}$. In  Supp. Fig.~\ref{fig:SingleShotResolutionCriterion}(a), the clustering quality is shown as function of the number of simulated signal photons  $\tilde{N}_\textrm{ph}$. The three curves show data for different ROI search regions, labeled by their respective start times. %Further, we limit the ROI search interval to the 20~ns after the start time, thereby enforcing that in each of the three curves, the algorithm operates on a different spectral feature. 
For each case, the necessary number of photons according to the single-shot criterion in Eq. (3) is indicated by the corresponding dashed vertical line. As expected, the clustering quality generally increases with rising signal photon number $\tilde{N}_\textrm{ph}$.  Supp. Fig.~\ref{fig:SingleShotResolutionCriterion}(b) shows the resulting models for different values of $\tilde{N}_\textrm{ph}$ higher than the resolution criterion. It can be seen that within each class, the models for different $\tilde{N}_\textrm{ph}$ agree well with each other, such that the criterion in \cref{eq:singleShot_N} indeed acts as  a conservative estimate of the photon number needed to resolve the feature. For the analysis from 3\,ns onward (top panel) it can further be seen that the clustering works at early times while the statistics is not sufficient to properly resolve later temporal features.

\section*{} % workaround to have following section title properly be displayed

\section{\label{simulated}Single-shot sorting-analysis of simulated data}
In order to verify the single-shot sorting approach also for temporal  features which are not accessible from the experimental data, we numerically simulated experimental data with higher mean photon number, following the experimentally observed photon-number distribution. For this, we fitted a negative binomial distribution to the estimated photon-number distribution of the two measurements separately, resulting in a mode number $M=1.14\pm0.11$ ($M=1.18\pm0.06$) and a mean signal photon number $\mu=218\pm8$ ($\mu=261\pm7$) for measurement A (B). Note that those mean signal photon numbers differ from the stated 101 photons/$(\mathrm{pulse}\cdot \gamma_0)$ delivered by the XFEL in Appendix~\ref{excitation} due to the superradiant broadening and losses. Based on the single-shot resolution criterion Eq.~(\ref{eq:singleShot_N}), we  chose an increased  mean photon number $\mu'=20\mu$ sufficient for all three cases in Supp. Fig.~\ref{fig:SingleShotResolutionCriterion}(a). Using this photon number distribution, we simulated 100,000 APD traces using the approach explained in Appendix~\ref{criterion}, and applied the single-shot sorting algorithm. The resulting recovered time spectra are shown in Supp. Fig.~\ref{fig:AnanlysisOfSimulatedData}. The shaded regions indicate the three ROIs over which the algorithm analyzed the shots, respectively. In all cases, we find that the analysis results agree well with the black dashed true time spectra. This confirms that the single-shot sorting algorithms can be successfully applied to general temporal features, provided that the recorded mean signal photon number is sufficiently high, as quantified by the resolution criterion.

\bibliographystyle{apsrev4-2-custom}
\bibliography{library}

\vskip 2cm

\vskip 0.5cm
\noindent
\textbf{Acknowledgements.}\\
We acknowledge European XFEL in Schenefeld, Germany, for provision of X-ray free-electron laser beamtime at MID under proposal number 3334 and would like to thank the staff for their assistance, in particular W. Jo, J. Möller, A. Parenti and J. Wrigley.

\clearpage

\setcounter{figure}{0}
\renewcommand{\figurename}{Supplemental Fig.}
\renewcommand{\thefigure}{S\arabic{figure}}

\begin{figure*}[t]
    \centering
    \includegraphics[width=15cm]{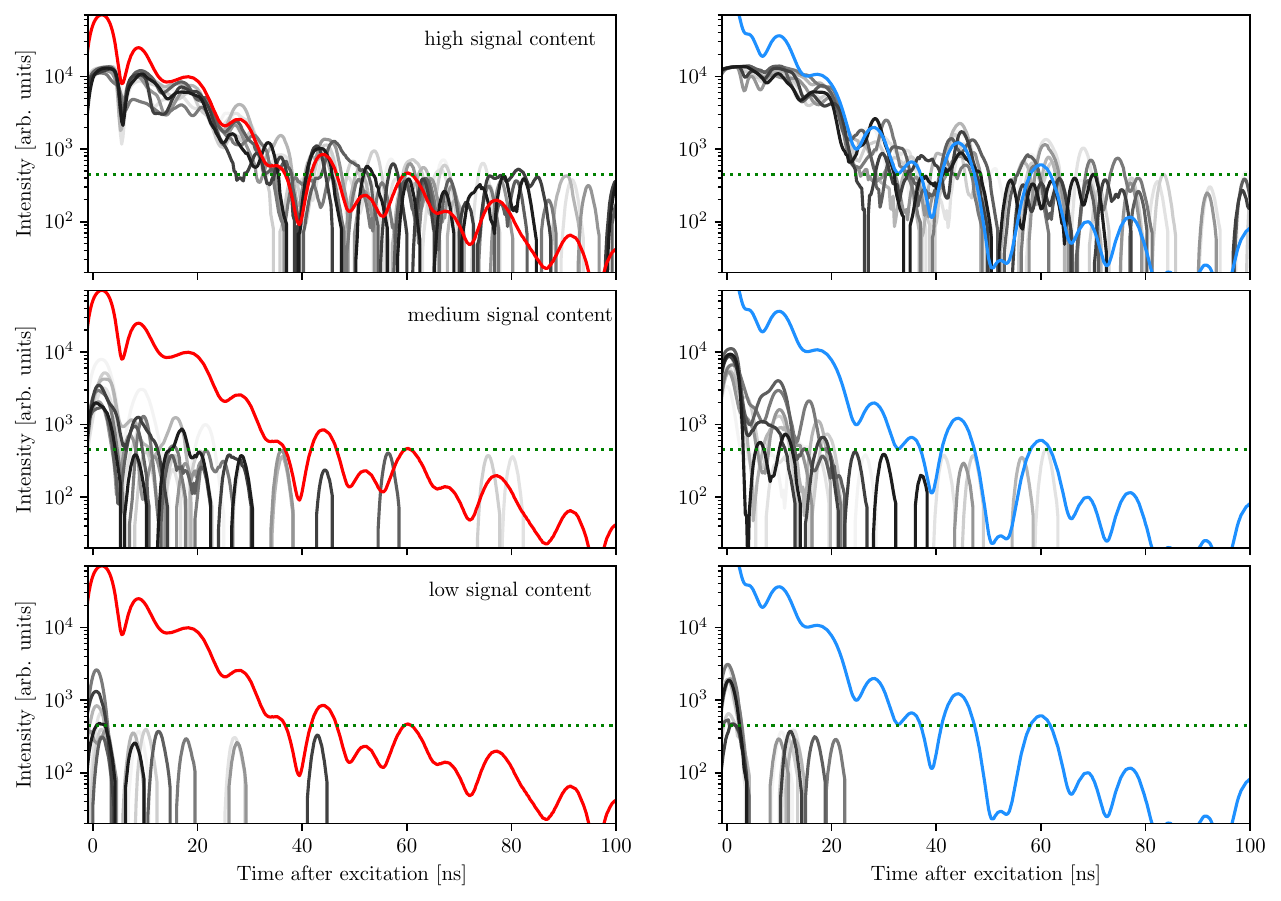}
    \caption{{ Example single-shot time-domain data.} For measurement A (left column) and B (right),  ten single-shot examples with high (top row), medium (center) and low (bottom) signal content, respectively, are shown in different shades of gray. The selected traces are the ones with highest signal content overall (high), of the lower half of the shots according to the signal content (medium) and of the lowest 10\% of the shots according to the signal content (low). In red (blue) the averaged data belonging to measurement A (B) are displayed. The green dashed line indicates the average height of the APD signal for individual recorded photons.}
    \label{fig:exampleTraces}
\end{figure*}

\begin{figure*}
    \centering
    \includegraphics[width=11cm]{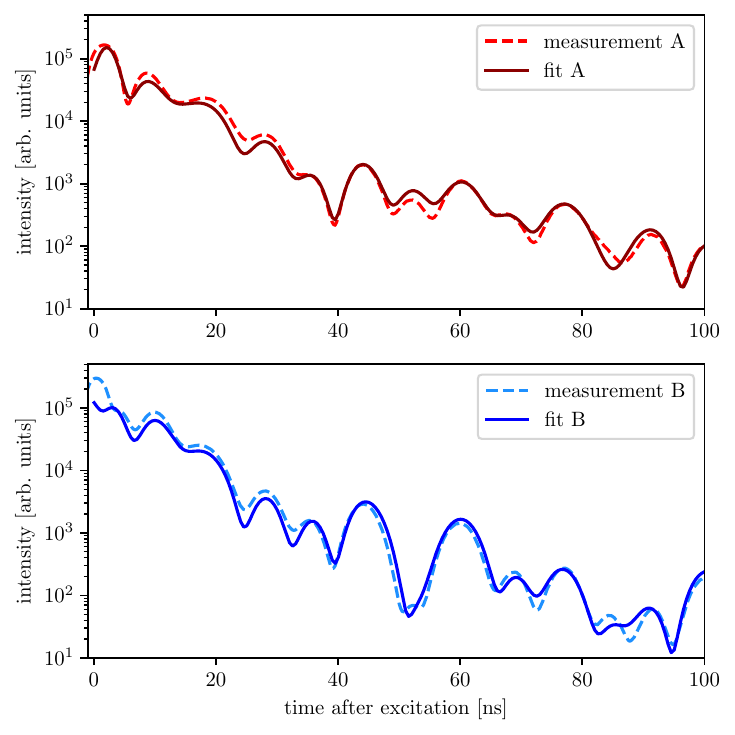}
    \caption{{ Raw data of the two measurements with theory fits.} The upper panel shows an average over all shots of dataset \rA, the lower one the corresponding results of dataset \rB. In both panels, the dashed line shows the experimental measurement, while the solid line shows the corresponding theory fit. Details are given in the Appendix~\ref{sample}. }
    \label{fig:rawDataFit}
\end{figure*}

\begin{figure*}
    \centering
    % EDITED new figure with new xlabel
    \includegraphics[width=10cm]{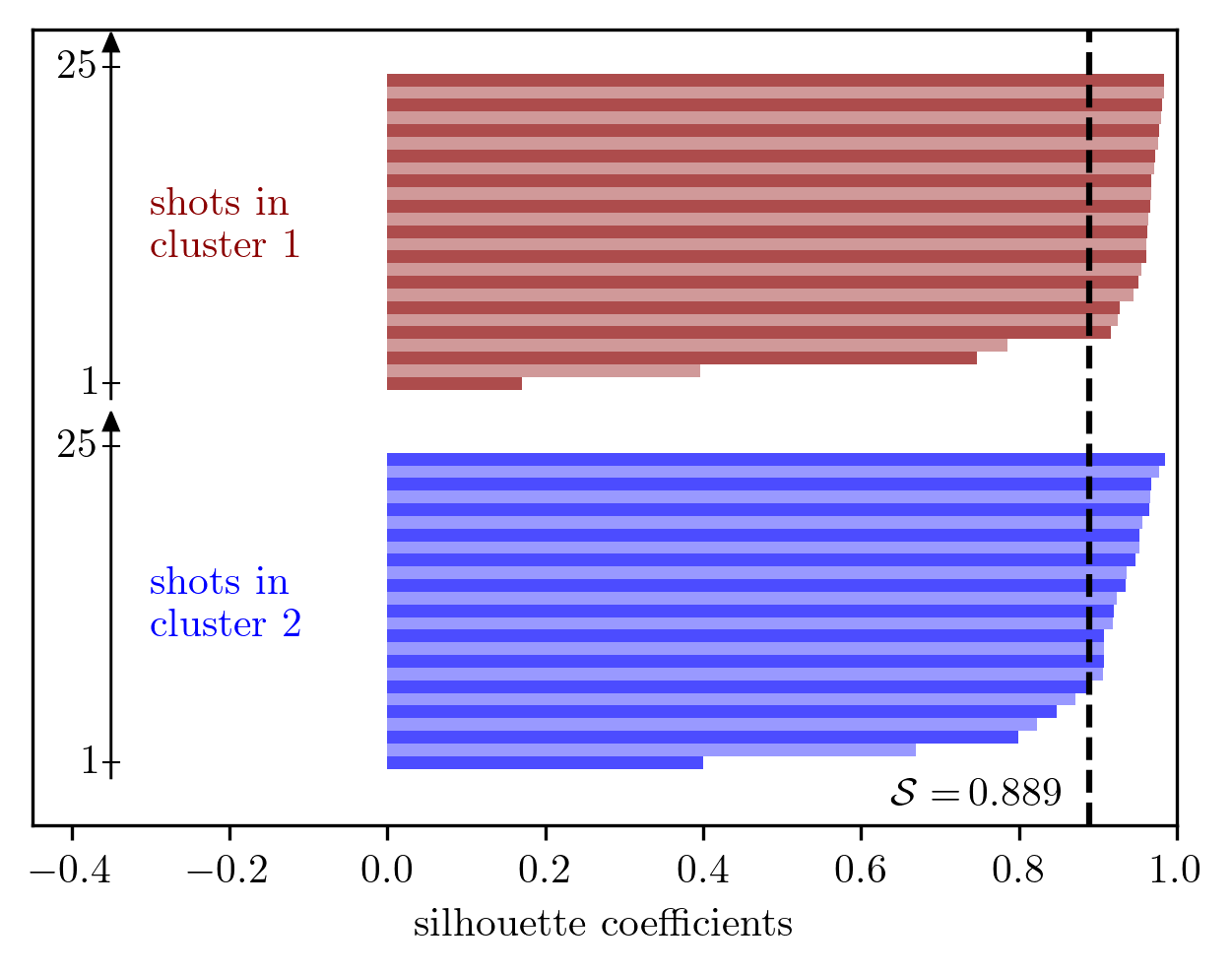}
    \caption{{ Individual silhouette coefficients for the shots used in the model building.} The figure shows the individual silhouette scores of the $N_\textrm{hs}=50$ shots of the experiment used for the model-building. The ROI for comparing different shots ranges from 3\,ns to 10\,ns. Each bar represents one shot with the colors indicating to which cluster the respective shot belongs. There are 9 shots in cluster 1, and 11 in cluster 2. The mean clustering quality $\mathcal{S}=0.889$ is represented by the dashed black line.}
    \label{fig:silhouette}
\end{figure*}

 \begin{figure*}
    \centering
    \includegraphics[width=15cm]{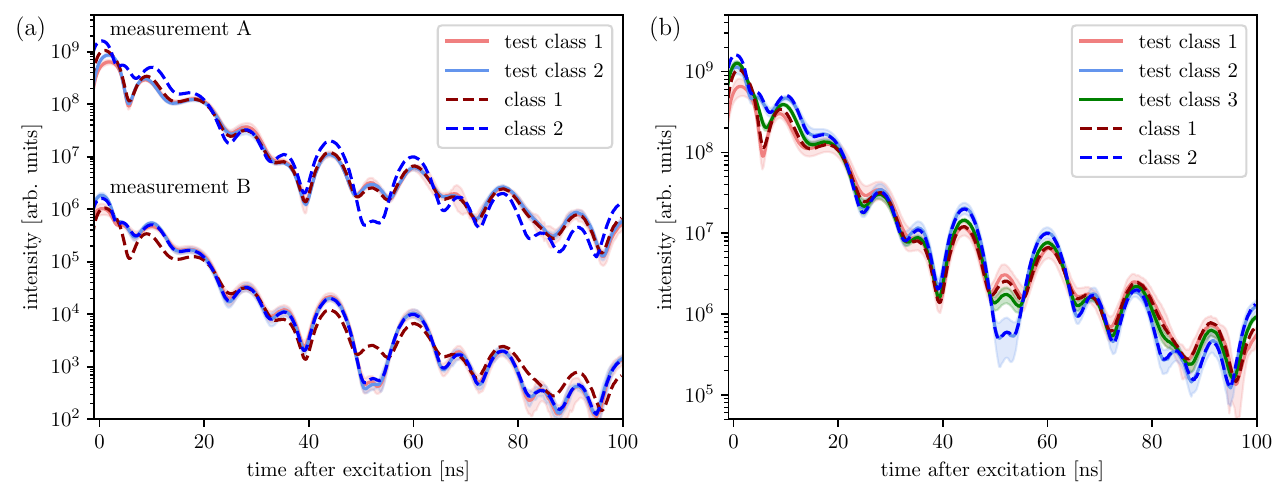}
    \caption{{ Stability of analysis}. (a) In the top part of the panel, the analysis is only performed on the shots from measurement A. The lines ``test class 1/2'' show the time-domain data of the two reconstructed classes. For comparison, the dashed lines ("class 1" and "class 2") represent the analysis results of the full dataset from the main text in Fig.~\ref{fig:spectra-datadriven}. It can be seen that for the dataset restricted to measurement A, the two recovered classes agree with each other within their uncertainty, as expected. Furthermore, they agree well with "class 1", which corresponds to measurement A. For completeness, the same analysis is performed with the data only from measurement B and shown in the lower part of the panel. For better visibility, the results for data for the two measurement settings are shifted  along the y-axis such that they can be better distinguished from each other. 
    (b) Results of the sorting analysis applied to the full dataset comprising measurement settings A and B, but with clustering into three classes. The results of these three test classes are shown as solid lines. The time-domain data is scaled for better comparability by fitting an overall scaling factor in the time range from 0ns to 100ns. The uncertainty bands of the different sets are calculated by a statistical analysis as explained in Appendix~\ref{stability}. 
    As a reference, also the results of the  sorting analysis from the main text in Fig.~\ref{fig:spectra-datadriven} are displayed in addition (dashed lines "class 1" and "class 2"). }
    \label{fig:stabilityAnalysis}
\end{figure*}

% EDITED: new figure
 \begin{figure*}
    \centering
    \includegraphics[width=15cm]{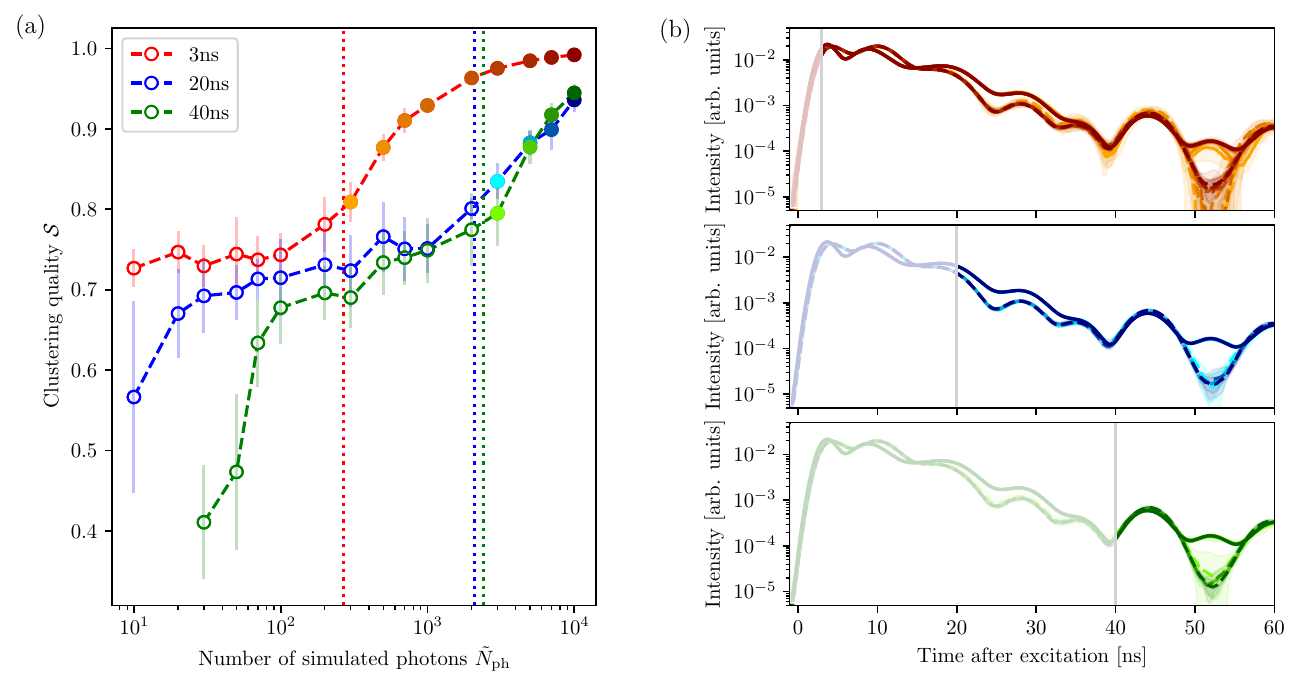}
    \caption{{ Single-shot resolution criterion}. (a) Clustering quality $\mathcal{S}$ as function of the number of simulated signal photons $\tilde{N}_\textrm{ph}$. This analysis uses data numerically simulated for a fixed photon number using the experimentally recorded spectra as the probability distribution. The different colors represent different start times of the analysis ROI, thereby enforcing that the clustering analysis operates on different spectral features. For each feature, the necessary number of photons according to the single-shot criterion in Eq.~(3) in the main text is indicated by the corresponding dashed vertical line. Empty (filled) circles represent results below (above) this threshold. 
    (b) Models resulting from the clustering analysis for the different $\tilde{N}_\textrm{ph}$ above the threshold. The three panels correspond to the different start times in (a) and the excluded time region is indicated by the shading and vertical line. The colors correspond to the colors of the filled circles in (a). The solid and dashed lines represent the two different classes, respectively. More details are given in Appendix~\ref{criterion}. 
    }
    \label{fig:SingleShotResolutionCriterion}
\end{figure*}

% EDITED: new figure
 \begin{figure*}
    \centering
    \includegraphics[width=10cm]{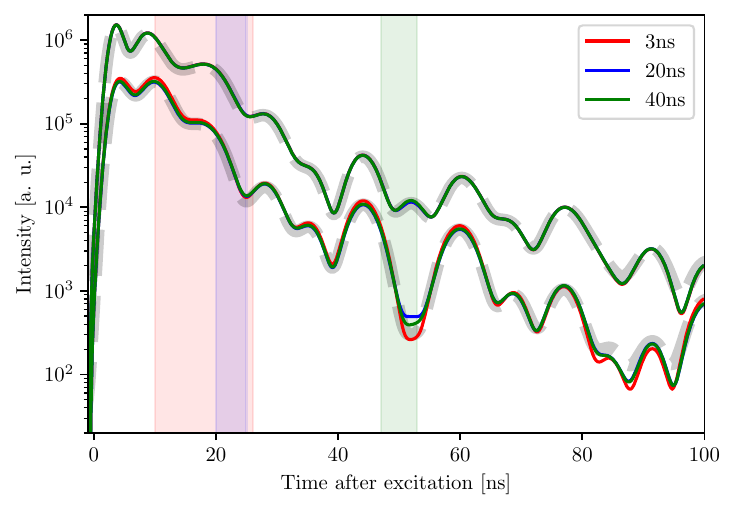}
    \caption{{ Single-shot sorting-analysis of simulated data}. The figure shows results of the single-sorting analysis on a simulated dataset drawn from the two experimentally recorded time-dependent intensities forming the different dynamics classes. The dataset follows the photon number statistics as observed in the experiment, except for a $\mu'/\mu =20$ times higher mean photon number representing an increased incident XFEL photon flux. The gray dashed lines show the simulated true spectra obtained by averaging over the respective datasets for the two classes. The colored lines represent the results from the single-shot analysis applied to the simulated data, with different initial time intervals excluded. This forces the algorithm to analyze different temporal features, leading to different respective analysis ROIs indicated by the shaded areas. More details are given in Appendix~\ref{simulated}. The reconstructed time-domain datasets are scaled with respect to the actual time-domain datasets for better comparability by fitting an overall scaling factor to the interval from 0ns to 100ns. For better visibility, the two cases are shifted relative to each other in the logarithmic plot by a factor of 5.
    }
    \label{fig:AnanlysisOfSimulatedData}
\end{figure*}

\end{document}